\documentclass[12pt]{iopart}

\usepackage{iopams}

\newcommand{\prt}{\partial}

\newcommand{\la}{\lambda}

\begin{document}

\title[On periodic solutions for the Manakov system]
{On periodic solutions and their modulations for the Manakov system}

\author{A M Kamchatnov}

\address{Institute of Spectroscopy, Russian Academy of Sciences,
Moscow, Troitsk, 142190, Russia}
\ead{kamch@isan.troitsk.ru}

\begin{abstract}
Periodic solutions of the Manakov system are studied with the aim to establish links between
recent results of Wright (Physica D 2013 {\bf 264}   1) and Kamchatnov (EPL 2013 \textbf{103}  60003),
where the one-phase solutions have been obtained by different methods and with different parameterizations.
Two types of waves, which in the context of dynamics of two-component Bose-Einstein condensates
can be called {\it density} and {\it polarization} waves, are determined as particular solutions
of the Manakov system. Whitham equations describing modulations of these two types of waves are
obtained.
\end{abstract}


\maketitle

\section{Introduction}

Periodic solutions of nonlinear wave equations play an important role in the theory of dispersive
shock waves and in other their applications to physics. If a nonlinear wave equation is completely
integrable by the inverse scattering transform approach, then periodic and quasi-periodic solutions of this equation
can be obtained by the powerful finite-gap integration method (see, e.g., \cite{belokolos-1994}
and references therein). In particular, the complete integrability of a very important for applications
nonlinear Schr\"odinger (NLS) equation was discovered in \cite{zs-71} and periodic solutions of
this equation were obtained by the finite-gap method in \cite{ik-76,ma-81,tc-88}. The solutions
found in this way do not have convenient enough form because an additional constraint is imposed upon the
variables used in the method. This constraint singles out the {\it real} solutions from the general
finite-gap formulae and it is difficult to satisfy this additional condition in the multi-phase quasi-periodic case.
As was shown in \cite{kamch-90}, the additional constraint can be resolved explicitly to determine
the loci of the auxiliary spectral variables used in the finite-gap method in the most important
for applications case of the single-phase periodic solutions of the NLS equation and later this
technique was extended to many other single-phase solutions of integrable equations \cite{kamch-97,kamch-2000}
which belong to the Ablowitz-Kaup-Newell-Segur (AKNS) scheme \cite{AKNS}.

There are integrable equations which do not belong to the AKNS scheme and the method of \cite{kamch-90}
cannot be applied to them directly. One of the most important such equations is the Manakov system
\cite{manakov} representing two coupled NLS equations and having applications to nonlinear optics and
Bose-Einstein condensates (BEC) physics. Some particular solutions of the Manakov system have
been studied in \cite{pp-99,ceek-2000,eek-2000} by introducing a special {\it ansatz} for the form
of the solution so that the problem is reduced to the AKNS-type scheme. The finite-gap integration
method was applied to the Manakov system by Shin \cite{shin-2003}, however, the reality conditions were
not resolved in this paper. Wright has recently succeeded \cite{wright-13} in extension of the method
of \cite{kamch-90} to the Manakov system so that the reality conditions are completely characterized
and the general formulae for the single-phase solution are obtained. It is also shown that the
Manakov soliton solution can be recovered in the corresponding limit. A similar solution was
obtained in \cite{kamch-13} by a direct method based on introduction of new dependent variables
whose form was motivated by physical interpretation of the Manakov system as the Gross-Pitaevskii
equations describing dynamics of the two-component BECs. In this context it is natural to
distinguish the in-phase and counter-phase motions of the components, and the new variables
permit one to describe separately these two modes of BEC dynamics. As a result, the periodic
solutions corresponding to the {\it density} (in-phase) and the {\it polarization} (counter-phase)
waves as well as to their combination were obtained. Such a distinction between two kinds of motion
has not been considered in \cite{wright-13} and the aim of this paper is to show that the solutions
studied in detail in \cite{kamch-13} can be obtained in framework of the scheme developed by
Wright.

The advantage of the finite-gap integration method is that the periodic solutions are parameterized
by the parameters which play the role of the Riemann invariants in the Whitham theory for
slowly modulated wavetrains \cite{whitham}. This property of the method was first demonstrated
for the Korteweg-de Vries equation in \cite{FFM-80} and it was extended to the NLS equation in
\cite{FL-86,pavlov-87}. A simple method of derivation of the Whitham equations for single-phase
wavetrains in framework of the finite-gap approach was suggested in \cite{kamch-90b,kamch-94}
and it was applied to many integrable equations in \cite{kamch-97,kamch-2000}. Wright supposed
that similar parameters (namely, branch points of the trigonal spectral curve which in the Manakov case
replaces the elliptic spectral curve appearing in the NLS case) play the same role in the Whitham
theory for the Manakov system and confirmed this supposition by a nontrivial example of the
dispersionless limit of this system \cite{wright-95}. In Section 3 of the present paper we consider
the Whitham equations for modulations of purely density waves and show that a simple
extension of the method developed in \cite{kamch-90b,kamch-94} leads to equations coinciding with
those \cite{FL-86,pavlov-87} for the NLS case. This result is physically natural---for synchronized
in-phase motion of two components difference between them disappears and their dynamics reduces
to the one-component one. However, the modulations of the polarization waves are not described by
the same parameters (i.e., by the branch points of the spectral curve) and the corresponding
Whitham equations for a purely polarization wave are obtained by the direct Whitham method of
averaging of the conservation laws. These Whitham equations can be transformed to the Riemann diagonal form
and they can be integrated the standard method of characteristics. The results obtained
for these two particular types of the periodic solutions of the Manakov system can be applied to
relevant problems of BEC dynamics.

\section{Periodic solutions}

In this section we describe briefly the main results of Refs.~\cite{wright-13} and \cite{kamch-13} with the goal
to establish links between these two forms of periodic solutions of the Manakov system.
We shall consider the Manakov equations in the form
\begin{equation}\label{eq1}
i \partial_t \psi_{\pm} + \frac12 \partial_{xx}^2\psi_{\pm} - \left(|\psi_\pm|^2+|\psi_{\mp}|^2\right) \psi_\pm =0
\end{equation}
corresponding to the repulsive interaction between atoms in the BEC physical context (for
definiteness we shall use throughout this interpretation of the Manakov system).
At first we shall present the solution in the form obtained by the direct integration of this system.

\subsection{Direct method}

The direct method is based on the introduction of such new dependent variables that the functions $\psi_{\pm}$
are represented by the formulae \cite{ktu-2005}
\begin{equation}\label{eq2}
    \left(
            \begin{array}{c}
              \psi_+ \\
              \psi_- \\
            \end{array}
          \right)=
          \sqrt{\rho}\, e^{i\Phi/2}
          \left(
            \begin{array}{c}
              \cos\frac{\theta}2\,e^{-i\phi/2} \\
              \sin\frac{\theta}2\,e^{i\phi/2}  \\
            \end{array}
          \right).
\end{equation}
Here $\rho(x,t)=\rho_++\rho_-=|\psi_+|^2+|\psi_-|^2$ denotes the total density of the
condensate and the angle $\theta(x,t)$ is the variable describing the
relative density of two components ($\cos\theta=(|\psi_+|^2-|\psi_-|^2)/\rho$);
$\Phi(x,t)$ has the meaning of the velocity potential of the
in-phase motion and $\phi(x,t)$ is the potential of the relative (counter-phase) motion
of the components. Accordingly, their gradients characterize the in-phase and relative velocities
\begin{equation}\label{eq3}
    U(x,t)=\Phi_x,\quad v(x,t)=\phi_x.
\end{equation}
The variables $\rho(x,t),\,U(x,t)$ describe the in-phase motion of the components ({\it density} waves)
and the variables $\theta(x,t),\,v(x,t)$ describe their counter-phase motion ({\it polarization} waves).
Substitution of (\ref{eq2}) and (\ref{eq3}) into (\ref{eq1}) yields the equations for
the variables $\rho$, $U$, $\theta$, $v$, (see \cite{kklp-13}) which can be written for future
convenience in the form of conservation laws,
\begin{eqnarray}
    &\rho_t+\frac12[\rho(U-v\cos\theta)]_x=0,
    \label{eq4}\\
    &U_t+\left[\frac{\rho_x^2}{4\rho^2}-\frac{\rho_{xx}}{2\rho}-\frac{\cot\theta}{2\rho}(\rho\theta_x)_x
+\frac14(\theta_x^2+U^2+v^2)+2\rho\right]_x=0,
\label{eq5}\\
    &(\rho\cos\theta)_t+\frac12[\rho(U\cos\theta-v)]_x=0,
    \label{eq6}\\
    &v_t+\left[\frac12Uv-\frac1{2\rho\sin\theta}(\rho\theta_x)_x\right]_x=0.
    \label{eq7}
\end{eqnarray}

The single-phase solution corresponds to the {\it ansatz} that all variables depend on $\xi=x-Vt$
and the solution can be obtained in an explicit form under supposition that an integration constant
appearing after integration of Eq.~(\ref{eq7}) equals to zero what means that the chemical potentials
of both BEC components are equal to each other, that is
\begin{equation}\label{eq8a}
    \Phi=-2\chi t+\Phi_0(\xi),\qquad \phi=\phi(\xi).
\end{equation}
where $\chi$ is the common value of the chemical potentials of both components.
Then, as was shown in \cite{kamch-13}, the density $\rho(\xi)$ satisfies the equation
\begin{equation}\label{eq8}
    \rho_{\xi}^2=4\mathcal{R(\rho)},
\end{equation}
where
\begin{equation}\label{eq9}
    \mathcal{R}(\rho)=\rho^3-(2\chi+V^2)\rho^2+D\rho-(A^2+C^2)=\prod_{i=1}^3(\rho-\rho_i),
\end{equation}
we assume that $\rho_1\leq\rho_2\leq\rho_3$ and parameters $\chi$, $A$, $C$ and $D$ are the integration constants.
The solution of (\ref{eq9})
can be expressed in standard notation in terms of the Jacobi elliptic function,
\begin{equation}\label{eq10}
    \rho(\xi)=\rho_1+(\rho_2-\rho_1)\mathrm{sn}^2(\sqrt{\rho_3-\rho_1}\,(\xi+\xi_0),m),
\end{equation}
where $m=(\rho_2-\rho_1)/(\rho_3-\rho_1)$. Thus, $\rho$ oscillates in the interval
$0\leq\rho_1\leq\rho\leq\rho_2$.

When the density $\rho(\xi)$ is known, the angle $\theta(\xi)$ can be found by integration of
the equation
\begin{equation}\label{eq11}
   \pm \frac{\sin\theta d\theta}{\sqrt{(\cos\theta-\cos\theta_1)(\cos\theta_2-\cos\theta)}}=2R\frac{d\xi}{\rho(\xi)},
\end{equation}
where the parameters $R$, $\theta_1$, $\theta_2$ are related with previously introduced parameters by the relations
\begin{eqnarray}
&R^2\equiv\rho_1\rho_2\rho_3 =A^2+C^2,\quad\theta_1=\beta+\gamma,\quad\theta_2=\beta-\gamma,\label{eq12}\\
    & A=R\cos\gamma,\quad C=R\sin\gamma,\quad B=R\cos\beta,\label{eq13}
\end{eqnarray}
so that $\cos\theta$ oscillates in the interval $\cos\theta_1\leq\cos\theta\leq\cos\theta_2$. Equation (\ref{eq11})
can be easily integrated to give
\begin{equation}\label{eq14}
    \cos\theta(\xi)=\cos\theta_1\sin^2\frac{X(\xi)}2+\cos\theta_2\cos^2\frac{X(\xi)}2,
\end{equation}
where
\begin{equation}\label{eq15}
    X(\xi)=2R\int_{\xi_0}^{\xi}\frac{d\xi'}{\rho_1+(\rho_2-\rho_1)\mathrm{sn}^2(\sqrt{\rho_3-\rho_1}\,\xi',m)}+X_0,
\end{equation}
$X_0$ is an integration constant.

At last, $U$ and $v$ are given by the formulae
\begin{equation}\label{eq16}
\eqalign{
    U(\xi)&=2V+\frac{2R}{\rho(\xi)}\frac{\cos\beta\cos\theta(\xi)-\cos\gamma}{\sin^2\theta(\xi)},\\
    v(\xi)&=\frac{2R}{\rho(\xi)}\frac{\cos\beta-\cos\gamma\cos\theta(\xi)}{\sin^2\theta(\xi)},}
\end{equation}
and their integration yields the phases $\Phi=-2\chi t+\int Udx$ and $\phi=\int vdx$.

Two particular solution are of special interest. If $\gamma=0$, then $\theta=\beta=\mathrm{const}$ and $v=0$,
that is there is no relative motion of the components. In particular, if $\rho_2=\rho_3=\rho_0$, then
the solution reduces to the standard Manakov dark soliton wave with
\begin{eqnarray}
    \rho(x,t)&=\rho_0\left\{1-\frac{1-\rho_1/\rho_0}{\cosh^2[\sqrt{\rho_0-\rho_1}(x-Vt)]}\right\},\label{eq17}\\
    U(x,t)&=2(V-\rho_0\sqrt{\rho_1}/\rho(x,t)).\label{eq18}
\end{eqnarray}
If soliton propagates through a quiescent condensate, that is $U\to0$, $\rho\to\rho_0$ as $|x|\to\infty$, then $V=\sqrt{\rho_1}$
and we reproduce a well-known one-soliton solution.  It does not depend on the parameter $\beta$,
because in this case there is no dynamical difference between the components and the motion reduces effectively
to a one-component dynamics: both components move in phase keeping the constant ratio of the densities.

The second important particular solution corresponds to the constant total density $\rho\equiv\rho_0$
when $\rho_1=\rho_2=\rho_0$, and hence $R=\rho_0\sqrt{\rho_3}$.
In such a wave the components exchange their places in the counter-phase relative motion called a pure
{\it polarization} dynamics. It is described by (\ref{eq14}) with
\begin{equation}\label{eq19}
    X(\xi)=2\sqrt{\rho_3}(\xi-\xi_0)+X_0.
\end{equation}
The total flux of atoms $v_+\rho_++v_-\rho_-=\rho_0(U-v\cos\theta)/2=\rho_0(V-\sqrt{\rho_3}\cos\gamma)$
vanishes when $V=\sqrt{\rho_3}\cos\gamma$. According to Eq.~(\ref{eq19}), the wave number of the polarization
wave equals to $k=2\sqrt{\rho_3}$, hence $V$ is the phase velocity $V=\omega/k$ corresponding to the dispersion
relation of the polarization waves given by
\begin{equation}\label{eq20}
   \omega(k)=\frac12k^2\cos\gamma,\qquad V=\frac{\omega}k=\frac12k\cos\gamma.
\end{equation}

More general solutions which combine oscillations of the total density $\rho$ and of the polarization variable
$\theta$ were considered in \cite{kamch-13}.

\subsection{Finite-gap method}

In this section we shall reproduce the main results of Wright's paper \cite{wright-13}
with the goal to relate them with the results of the preceding section. For consistency, we shall slightly change
the notation of \cite{wright-13} and make some additions to Wright's results.

The finite-gap method is based on the possibility to represent the Manakov system (\ref{eq1}) as a
compatibility condition of two linear systems (see \cite{manakov})
\begin{equation}\label{eq21}
    \Psi_x=\mathbb{U}\Psi,\qquad \Psi_t=\mathbb{V}\Psi,
\end{equation}
that is from the condition
\begin{equation}\label{eq21a}
    \mathbb{U}_t-\mathbb{V}_x+[\mathbb{U},\mathbb{V}]=0,
\end{equation}
where $[\cdot,\cdot]$ is a commutator of matrices and
\begin{equation}\label{eq22}
    \mathbb{U}=\left(
                 \begin{array}{ccc}
                   F_1+F_2 & G_1 & G_2 \\
                   H_1 & -F_1 & G_3 \\
                   H_2 & H_3 & -F_2 \\
                 \end{array}
               \right),\qquad
               \mathbb{V}=\left(
                            \begin{array}{ccc}
                              A_1+A_2 & B_1 & B_2 \\
                              C_1 & -A_1 & B_3 \\
                              C_2 & C_3 & -A_2 \\
                            \end{array}
                          \right),
\end{equation}
\begin{equation}\label{eq23}
\eqalign{
    &F_1=F_2=-i\la,\qquad G_1=i\psi_+,\qquad G_2=i\psi_-,\qquad G_3=0,\\
    &H_1=-i\psi_+^*,\qquad H_2=-i\psi_-^*,\qquad H_3=0;}
\end{equation}
\begin{equation}\label{eq24}
    \eqalign{
    &A_1=-\frac32i\la^2-\frac12i|\psi_+|^2, \qquad A_2=-\frac32i\la^2-\frac12i|\psi_-|^2,\\
    &B_1=\frac32i\la\psi_+-\frac12\psi_{+,x},\qquad B_2=\frac32i\la\psi_--\frac12\psi_{-,x},
    \qquad B_3=\frac12i\psi_+^*\psi_-, \\
    &C_1=-\frac32i\la\psi_+^*-\frac12\psi_{+,x}^*,\qquad C_2=-\frac32i\la\psi_-^*-\frac12\psi_{-,x}^*,
    \qquad C_3=\frac12i\psi_+\psi_-^*.}
\end{equation}
Equations (\ref{eq21}) can be cast to the matrix form
\begin{equation}\label{eq25}
    \mathbb{W}_x=[\mathbb{U},\mathbb{W}],\qquad \mathbb{W}_t=[\mathbb{V},\mathbb{W}],
\end{equation}
where $\mathbb{W}$ can be built from the basis
solutions $\Psi_0$ of the systems (\ref{eq21}) (see, e.g., \cite{shin-2003}). The explicit formulae
expressing the matrix elements of $\mathbb{W}$ in terms of $\Psi_0$ are presented in Appendix A.
Let us define a characteristic polynomial of the matrix $\mathbb{W}$ by the equation
\begin{equation}\label{eq26}
    \mathcal{Q}(\la,w)=i\mathrm{det}[iw\mathbb{I}-\mathbb{W}]=w^3+a(\la)w+b(\la),
\end{equation}
where $\mathbb{I}$ is the identity matrix. Then from the fact that $\mathbb{W}$ satisfies the equations (\ref{eq25})
one can obtain the theorem that the coefficients of the polynomial $\mathcal{Q}$ do not depend
on $x$ and $t$. Since these coefficients are expressed in terms of the matrix elements of $\mathbb{U}$
and $\mathbb{V}$, that is in terms of $\psi_{\pm}$ and their $x$-derivatives, we arrive at the
conservation laws of the Manakov system (\ref{eq1}). Simple proof of this theorem is presented
for completeness in Appendix B.

Under supposition that the matrix elements of $\mathbb{W}$ are the second-degree polynomials in $\la$,
Wright obtains the solution
\begin{equation}\label{eq27}
    \mathbb{W}=\left(
                 \begin{array}{ccc}
                   f_1+f_2 & g_1 & g_2 \\
                   h_1 & -f_1 & g_3 \\
                   h_2 & h_3 & -f_2 \\
                 \end{array}
               \right),
\end{equation}
where
\begin{equation}\label{eq28}
\eqalign{
    f_1&=-i\la^2-\frac23iV\la-\frac13i|\psi_+|^2+\frac29i\chi,\\
    f_2&=-i\la^2-\frac23iV\la-\frac13i|\psi_-|^2+\frac29i\chi,\\
    g_1&=i\psi_+(\la-\mu_1),\qquad g_2=i\psi_-(\la-\mu_2),\qquad g_3=\frac13i\psi_+^*\psi_-,\\
    h_1&=-i\psi_+^*(\la-\mu_1^*),\qquad h_2=-i\psi_-^*(\la-\mu_2^*),\qquad h_3=-\frac13i\psi_+\psi_-^*,}
\end{equation}
$V$ and $\chi$ are constant parameters, and the coefficients of these polynomials satisfy the equations
\begin{equation}\label{eq29}
\eqalign{
    \psi_{+,x}&=3i\mu_1\psi_++2iV\psi_+,\qquad \psi_{-,x}=3i\mu_2\psi_++2iV\psi_-,\\
    \psi_{+,t}&=-V\psi_{+,x}-i\chi\psi_+,\qquad \psi_{-,t}=-V\psi_{-,x}-i\chi\psi_-,\\
    \mu_{1,x}&=-3i\mu_1^2-2iV\mu_1-\frac23i(|\psi_+|^2+|\psi_-|^2)+\frac23i\chi,\\
    \mu_{2,x}&=-3i\mu_2^2-2iV\mu_2-\frac23i(|\psi_+|^2+|\psi_-|^2)+\frac23i\chi,\\
    \mu_{1,t}&=-V\mu_{1,x},\qquad \mu_{2,t}=-V\mu_{2,x}.}
\end{equation}
From these equations one can obtain
\begin{equation}\label{eq30}
    \left(|\psi_+|^2\right)_x=3i|\psi_+|^2(\mu_1-\mu_1^*),\qquad \left(|\psi_-|^2\right)_x=3i|\psi_-|^2(\mu_2-\mu_2^*).
\end{equation}
One can see that the variables $\mu_1,\,\mu_2$ as well as $\rho_{\pm}=|\psi_{\pm}|^2$ depend on $\xi=v-Vt$ only,
that is $V$ denotes the phase velocity of the wave and $\chi$ plays the role of the chemical potential in the
BEC context. In this particular case of the second degree polynomials (\ref{eq28}) in $\la$ the coefficients
of the polynomial $\mathcal{Q}$ take the form
\begin{equation}\label{eq31}
\eqalign{
    a(\widetilde{\la})=&-3\left[\widetilde{\la}^2-\frac19(V^2+2\chi)\right]^2-\left(\widetilde{\la}-\frac13V\right)S-I_3,\\
    b(\widetilde{\la})=&2\left[\widetilde{\la}^2-\frac19(V^2+2\chi)\right]^3 \\
    &+\left[\left(\widetilde{\la}-\frac13V\right)S+I_3\right]
    \left[\widetilde{\la}^2-\frac19(V^2+2\chi)\right]-\frac13I_4,}
\end{equation}
where $\widetilde{\la}=\la-V/3$, $S=I_1+I_2$, and $I_1,\,I_2,\,I_3,\,I_4$ are the integrals introduced by Wright:
\begin{equation}\label{eq32}
\eqalign{
    &I_1=|\psi_+|^2(\mu_1^*+\mu_1+2V/3),\qquad I_2=|\psi_-|^2(\mu_2^*+\mu_2+2V/3),\\
    &I_3=\frac19(|\psi_+|^2+|\psi_-|^2)^2-|\psi_+|^2|\mu_1|^2-|\psi_-|^2|\mu_2|^2-\frac29\chi(|\psi_+|^2+|\psi_-|^2),\\
    &I_4=|\psi_+|^2|\psi_-|^2|\mu_1-\mu_2|^2.}
\end{equation}
It is important to notice that the polynomial $\mathcal{Q}$ depends on the sum $S$ of integrals $I_1$ and $I_2$ only
rather than on them separately.

If we introduce the parameter $\widetilde{\beta}=(V^2+2\chi)/9$ and make a substitution $w=\widetilde{\la}^2-\widetilde{\beta}+z$
in $\mathcal{Q}=w^3+a(\la)w+b(\la)$, then we obtain $\mathcal{Q}$ in new variables
$$
\mathcal{Q}=z^3+3(\widetilde{\la}^2-\widetilde{\beta})z^2-[(\widetilde{\la}-V/3)S+I_3]z-\frac13I_3.
$$
One can notice that this is a quadratic in $\widetilde{\la}z$ polynomial so that we arrive at the identity
\begin{equation}\label{eq33}
    \mathcal{Q}=w^3+a(\la)w+b(\la)\equiv \frac1{27}\left[y^2+\mathcal{R}(\nu)\right],
\end{equation}
where $w,\la$ are related with the variables $y=3z,\,\nu$ by the equations
\begin{equation}\label{eq34}
    \la=\frac1{\nu}\left(\frac{y}3+\frac{S}2\right)-\frac{V}3,\qquad
    w=\frac1{\nu^2}\left(\frac{y}3+\frac{S}2\right)^2-\frac19(V^2+2\chi)+\frac{\nu}3,
\end{equation}
and
\begin{equation}\label{eq35}
    \mathcal{R}(\nu)=\nu^3-(2\chi+V^2)\nu^2+9\left(\frac13VS-I_3\right)\nu-\frac94(S^2+4I_4).
\end{equation}

Using the integrals (\ref{eq32}), Wright has found useful expressions for the parameters
$\mu_1,\,\mu_2$ which we write in the form
\begin{equation}\label{eq36}
\eqalign{
    \mu_1&=\frac12\left(\frac{I_1}{|\psi_+|^2}-\frac{2V}3\right)\pm\frac{i\sqrt{\mathcal{S}}}{2\rho|\psi_+|^2}
    +\frac{i\sqrt{\mathcal{R}(\rho)}}{3\rho},\\
    \mu_2&=\frac12\left(\frac{I_2}{|\psi_-|^2}-\frac{2V}3\right)\mp\frac{i\sqrt{\mathcal{S}}}{2\rho|\psi_-|^2}
    +\frac{i\sqrt{\mathcal{R}(\rho)}}{3\rho},}
\end{equation}
where the function $\mathcal{R}$ is the same as in (\ref{eq35}), $\rho=|\psi_+|^2+|\psi_-|^2$ is the
total density, and $\mathcal{S}$ is defined by the expression
\begin{equation}\label{eq37}
    \mathcal{S}=4|\psi_+|^2|\psi_-|^2I_4-(I_1|\psi_-|^2-I_2|\psi_+|^2)^2.
\end{equation}
Substitution of (\ref{eq36}) in the sum of (\ref{eq30}) yields the equation for $\rho$:
\begin{equation}\label{eq38}
    \rho_x^2=4\mathcal{R}(\rho).
\end{equation}
If we introduce the variable $\theta$ characterizing the relative density of the components according to
\begin{equation}\label{eq39}
    1+\cos\theta=2\cos^2\frac{\theta}2=\frac{2|\psi_+|^2}{\rho},
\end{equation}
then we can find equation for $\theta$ with the use of (\ref{eq30}) and (\ref{eq38})
\begin{equation}\label{eq40}
\eqalign{
    &\left(\frac{d\cos\theta}{dx}\right)^2=\frac{36}{\rho^4}\,\mathcal{S}\\
    &=-\frac9{\rho^2}\left[(S^2+4I_4)\cos^2\theta-2(I_1^2-I_2^2)\cos\theta+(I_1-I_2)^2-4I_4\right].}
\end{equation}
Since $\rho$ and $\theta$ depend on $\xi=x-Vt$ only, their $x$-derivatives coincide with $\xi$-derivatives.
Hence, the functions $\mathcal{R}$ and $\mathcal{S}$ govern the evolution of $\rho$ and $\theta$ in the
single-phase solution of the Manakov system.

Equations (\ref{eq38}) and (\ref{eq40}) can be identified with equations (\ref{eq8}) and
(\ref{eq11}) of the preceding Section provided
\begin{equation}\label{eq41}
\eqalign{
    R^2=\frac94(S^2+I_4),\qquad A=\frac32S=R\cos\gamma,\qquad
    C=3\sqrt{I_4}=R\sin\gamma,\\
    \cos\beta=\frac{I_1-I_2}{\sqrt{S^2+4I_4}},\qquad \cos\gamma=\frac{S}{\sqrt{S^2+4I_4}}.}
\end{equation}
From the last two equations with account of $S=I_1+I_2$ we find
\begin{equation}\label{eq42}
    I_1=\frac{R}3(\cos\gamma+\cos\beta),\qquad I_2=\frac{R}3(\cos\gamma-\cos\beta)
\end{equation}
and
\begin{equation}\label{eq43}
    \mathcal{S}=\frac{R^2\rho^2}9(\cos\theta-cos\theta_1)(\cos\theta_2-\cos\theta),
\end{equation}
where $\theta_1=\beta+\gamma$, $\theta_2=\beta-\gamma$ (see (\ref{eq12})).
Then velocities $v_{\pm}=(U\mp v)/2$ of two components can be written as
\begin{equation}\label{eq44}
    v_+=V-\frac{3I_1}{2\rho\cos^2(\theta/3)},\qquad v_-=V-\frac{3I_2}{2\rho\sin^2(\theta/3)}.
\end{equation}
With the use of new parameters the equations (\ref{eq36}) can be written as
\begin{equation}\label{eq45}
\eqalign{
    \mu_1&=-\frac{v_+}3\pm\frac{i\sqrt{\mathcal{S}}}{2\rho^2\cos^2(\theta/2)}+\frac{i\sqrt{\mathcal{R}}}{3\rho},
    \\
    \mu_2&=-\frac{v_-}3\mp\frac{i\sqrt{\mathcal{S}}}{2\rho^2\sin^2(\theta/2)}+\frac{i\sqrt{\mathcal{R}}}{3\rho}.}
\end{equation}
Here $\mathcal{R}=\mathcal{R}(\rho)$, $\mathcal{S}=\mathcal{S}(\rho,\cos\theta)$ and both $\rho$ and $\cos\theta$ are
functions of $\xi=x-Vt$ given by (\ref{eq10}) and (\ref{eq14}), correspondingly.
Hence, equations (\ref{eq45}) define paths of the auxiliary spectrum points $\mu_{1,2}$
in the complex plane $\la$ corresponding to the single-phase evolution of $\psi_{\pm}$ according to the
Manakov system.

The single-phase solution is parameterized here by the integrals $I_1,\,I_2,\,I_3,\,I_4$ and by the velocity $V$
and the chemical potential $\chi$.
However, in the finite-gap integration method a special role is played by the parameters defining the {\it spectral
curve}. In our case such a curve is given by the equation
\begin{equation}\label{eq46}
    \mathcal{Q}(\la,w)=w^3+a(\la)w+b(\la)=0
\end{equation}
which defines the dependence of the complex variable $w$ on the complex variable $\la$. As we know, this equations
has constant coefficients expressed in terms of $S=I_1+I_2$, $I_3$, $I_4$, $V$ and $\chi$ (see (\ref{eq31})). Thus,
there are only five independent parameters which cannot parameterize completely the single-phase solution. Another
set of convenient parameters is presented by the branching points of the curve $w=w(\la)$ defined by (\ref{eq46}).
These points are determined by the condition that the discriminant of (\ref{eq46}) vanishes. The discriminant
can be defined as
\begin{equation}\label{eq47}
    \Delta(\la)=-\frac19[4a^3(\la)+27b^2(\la)]
\end{equation}
and it equals to
\begin{equation}\label{eq48}
\eqalign{
    &\Delta(\la)=(S\la+I_3)^2\left[\left(\la^2+\frac23V\la-\frac29\chi\right)^2+\frac49(S\la+I_3)\right]\\
    &+2\left(\la^2+\frac23V\la-\frac29\chi\right)\left[2\left(\la^2+\frac23V\la-\frac29\chi\right)^2
    +(S\la+I_3)\right]I_4-\frac13I_4^2.}
\end{equation}
This is sixth-degree polynomial in $\la$ and equation $\Delta(\la)=0$ has six roots $\la_i$, $i=1,\ldots,6,$
representing the branching points of the function $w(\la)$ in the complex $\la$-plane. They are not independent
of each other since they depend on five free parameters only. On the other hand, our initial parameters
$S=I_1+I_2$, $I_3$, $I_4$, $V$ and $\chi$ can be, in principle, expressed in terms of $\la_i$, $i=1,\ldots,6$,
and, hence, these branching points provide the most natural parametrization of the solution in the context of the finite-gap
integration method. In particular, using the first two terms of the expansion
$$
\Delta(\la)=(4I_4+S^2)\la^6+\frac23[3I_3S+2(6I_4+S^2)V]\la^5+\ldots
$$
we find the relation
\begin{equation}\label{eq49}
    V=-\frac{3(4I_4+S^2)}{4(6I_4+S^2)}\sum_{i=1}^6\la_i-\frac{3I_3S}{2(6I_4+S^2)}.
\end{equation}

As one can see from (\ref{eq48}), the situation simplifies drastically for $I_4=0$. Let us consider
this particular case in some detail. The discriminant reduces to
\begin{equation}\label{eq50}
    \Delta(\la)=(S\la+I_3)^2P_0(\la),
\end{equation}
where
\begin{equation}\label{eq51}
\eqalign{
    P_0(\la)&=\left(\la^2+\frac23V\la-\frac29\chi\right)^2+\frac49(S\la+I_3)\\
    &=\la^4+\frac43V\la^3+\frac49(V^2-\chi)\la^2+\frac4{27}(3S-2V\chi)\la+\frac49I_3+\frac4{81}\chi^2}
\end{equation}
is the forth-degree polynomial in $\la$ with zeroes $\la_{0i}$, $i=1,\ldots,4$. The doubled zero of the
discriminant equals to $\la_{05}=\la_{06}=-I_3/S$. Then (\ref{eq49}) reduces to
\begin{equation}\label{eq52}
    V=-\frac34\sum_{i=0}^4\la_{0i}
\end{equation}
and that is confirmed by the Vi\`ete formulae for (\ref{eq51}). Calculation of the Ferrari resolvent
(see \cite{Waerden} and \cite{kamch-90}) for the polynomial $P_0(\la)$ yields
\begin{equation}\label{eq53}
\eqalign{
    \mathcal{R}_0(\widetilde{\nu})=&\left(\frac49\right)^3\Big\{\left(\frac94\widetilde{\nu}\right)^3
    -(V^2+2\chi)\left(\frac94\widetilde{\nu}\right)^2\\
    &+9\left(\frac13SV-I_3\right)\left(\frac94\widetilde{\nu}\right) -\frac94S^2\Big\}.}
\end{equation}
This coincides (up to a constant numerical factor) with the polynomial (\ref{eq35}) where it is assumed
that $I_4=0$ and $\nu=(9/4)\widetilde{\nu}$. Consequently, we immediately obtain the relations between the
zeroes $\rho_{0i}$ of (\ref{eq35}) (with $I_4=0$) and the zeroes $\la_{0i}$ of the polynomial $P_0(\la)$,
\begin{equation}\label{eq54}
\eqalign{
    \rho_{01}&=(\la_{01}-\la_{02}-\la_{03}+\la_{04})^2/9,\\
    \rho_{02}&=(\la_{01}-\la_{02}+\la_{03}-\la_{04})^2/9,\\
    \rho_{03}&=(\la_{01}+\la_{02}-\la_{03}-\la_{04})^2/9}
\end{equation}
(see \cite{kamch-90,kamch-2000,wright-13}). Here the zeroes are ordered in such a way that for
$\la_{01}\leq\la_{02}\leq\la_{03}\leq\la_{04}$ we get $\rho_{01}\leq\rho_{02}\leq\rho_{03}$.
These formulae suggest that the case with $I_4=0$ reduces to the situation known for the theory
of periodic solutions of the one-component NLS equation \cite{kamch-90,kamch-2000} with the
polynomial $P_0(\la)$ playing the role of the polynomial appearing in the elliptic spectral
curve in the NLS finite-gap theory. This suppositions is supported by noticing that in the case
with $I_4=0$ the angle $\gamma$ defined by Eq.~(\ref{eq41}) vanishes that is in these periodic
solutions there is no variations of the relative densities of the components and dynamical
difference between them disappears. Formally we have $\theta=\theta_1=\theta_2=\beta=\mathrm{const}$
and consequently $\mathcal{S}\equiv0$. Then from (\ref{eq37}) we get
$$
\frac{I_1}{|\psi_+|^2}=\frac{I_2}{|\psi_-|^2}=\frac{I_1+I_2}{|\psi_+|^2+|\psi_-|^2}=\frac{S}{\rho},
$$
hence
\begin{equation}\label{eq55}
    \mu_1=\mu_2=\mu=\frac{S}{2\rho}-\frac{V}3+\frac{i\sqrt{\mathcal{R(\rho)}}}{3\rho}.
\end{equation}
Now the polynomial $P_0(\la)$ satisfies the identity (see \cite{kamch-90,kamch-2000})
\begin{equation}\label{eq56}
    P_0(\la)=f_0^2(\la)-\frac49\rho(\la-\mu)(\la-\mu^*),
\end{equation}
where
\begin{equation}\label{eq57}
    f_0(\la)=\la^2+\frac23V\la+\frac29(\rho-\chi)
\end{equation}
what equals to the value of $if_{1,2}$ defined by (\ref{eq28}) for the case $I_4=0$.
Consequently, equations for $\mu_{1,x}$ and $\mu_{2,x}$ in (\ref{eq29}) reduce to
\begin{equation}\label{eq58}
    \mu_x=-i\left(3\mu^2+2V\mu+\frac23(\rho-\chi)\right)=-3if_0(\mu)=-3i\sqrt{P_0(\mu)},
\end{equation}
that is to the well-known Dubrovin equation for a single-phase solution of the NLS equation
(see, e.g., \cite{kamch-90,kamch-97,kamch-2000}).

Thus, we have reproduced the main equations of the preceding Section obtained by a direct method in \cite{kamch-13}
by means of the finite-gap integration method developed by Wright \cite{wright-13} and have related the
finite-gap equations for the Manakov system with the finite-gap equations for the NLS equation in the
limit of density waves corresponding to $I_4=0$. The advantage of the finite-gap integration
method is that the branching points of the appearing in this method spectral curve provide the
Riemann invariants of the Whitham modulation equations. In the next Section we shall consider Whitham
equation for the particular cases of density and polarization waves.

\section{Modulation theory}

\subsection{Whitham equations for density waves}

At first let us consider the dispersionless limit when the dispersion effects are neglected that is all
higher order $x$-derivatives are omitted in the system (\ref{eq4})--(\ref{eq7}):
\begin{equation}\label{eq59}
\eqalign{
    &\rho_t+\frac12[\rho(U-v\cos\theta)]_x=0,\\
    &U_t+\left(\frac14U^2+\frac14v^2+2\rho\right)_x=0,\\
    &(\rho\cos\theta)_t+\frac12[\rho(U\cos\theta-v)]_x=0,\\
    &v_t+\frac12(Uv)_x=0.}
\end{equation}
If we introduce variables $r_W=\rho$, $s_W=\rho\cos\theta$, $k_W=U/2$, $d_W=-v/2$, then these equations
transform to the dispersionless limit of Manakov equations
\begin{equation}\label{eq60}
    \left(
      \begin{array}{c}
        k_W \\
        d_W \\
        r_W \\
        s_W \\
      \end{array}
    \right)_t+
    \left(
      \begin{array}{cccc}
        k_W & d_W & 1 & 0 \\
        d_W & k_W & 0 & 0 \\
        r_W & s_W & k_W & d_W \\
        s_W & r_W & d_W & k_W \\
      \end{array}
    \right)
    \left(
      \begin{array}{c}
        k_W \\
        d_W \\
        r_W \\
        s_W \\
      \end{array}
    \right)_x=0
\end{equation}
obtained by Wright \cite{wright-95}.

We are interested here in modulations of the density wave along which the conditions $v=0$ and $\theta=\beta=\mathrm{const}$
hold. Consequently, the first and the third equations of the system (\ref{eq59}) coincide with each other in this case
and the fourth equation is fulfilled identically. Thus, we remain with two equations for $\rho$ and $U$ only
which coincide with the Euler equations for polytropic gas flow with the ratio of specific heats equal to 2
(or to ``shallow water equations'')
\begin{equation}\label{eq61}
    \rho_t+(\rho \widetilde{U})_x=0,\qquad \widetilde{U}_t+\widetilde{U}\widetilde{U}_x+\rho_x=0,
\end{equation}
where $\widetilde{U}=U/2$.
Equations for this system in the Riemann invariant form are well known and can be written as
\begin{equation}\label{eq62}
    \la_{1,t}+\frac14(3\la_1+\la_2)\la_{1,x}=0,\qquad \la_{2,t}+\frac14(\la_1+3\la_2)\la_{2,x}=0,
\end{equation}
where
\begin{equation}\label{eq63}
    \la_1=-\frac13\left(\widetilde{U}-2\sqrt{\rho}\right),\qquad \la_2=-\frac13\left(\widetilde{U}+2\sqrt{\rho}\right)
\end{equation}
are the Riemann invariants.

We are interested in demonstration that these Riemann invariants can be obtained as the branching points
of the spectral curve, that is, in the density wave limit, as the zeroes of the polynomial $P_0(\la)$
defined in (\ref{eq51}). To this end, we express the coefficients of $P_0(\la)$ in terms of $\rho$ and $U$
assuming that the merging zeroes of $\mathcal{R(\nu)}$ are equal to $\rho_1=\rho_2=\rho$.
From (\ref{eq9}) and (\ref{eq16}) we get
$$
2\rho+\rho_3=2\chi+V^2,\qquad U=2(V-\sqrt{\rho_3}),
$$
consequently $\sqrt{\rho_3}=V-U/2$ and
\begin{equation}\label{eq64}
    \chi=\rho+\frac18U^2-\frac12UV.
\end{equation}
Then we have
\begin{equation*}
\frac94S^2=\rho^2\rho_3=\rho^2(V-U/2)^2,
\end{equation*}
and in taking the square root we have to choose such a sign that two of the resulting roots $\la_i$ do not
depend on $V$ according to the requirement that in the dispersionless limit the phase velocity of the wave
becomes irrelevant parameter. The final result will show that the appropriate choice of sign corresponds to
\begin{equation}\label{eq65}
    S=\frac29\rho(U/2-V).
\end{equation}
At last, $\mu=S/(2\rho)-V/3=(U/2-2V)/3$ and definition of $I_3$ in (\ref{eq32}) leads to
\begin{equation}\label{eq66}
    I_3=\frac19\rho^2-\mu^2\rho-\frac29\mu\rho=-\frac1{18}(U^2-6UV+8V^2+2\rho).
\end{equation}
Substitution of these values of the parameters into (\ref{eq51}) yields
\begin{equation}\label{eq67}
    P_0(\la)=\frac1{1296}(6\la-U+4V)^2(36\la^2+12U\la+U^2-16\rho)
\end{equation}
and the roots of this polynomial are equal to (to simplify the notation, in what follows we omit zero in
the index of $\la_{0i}$ for the case $I_4=0$)
\begin{equation}\label{eq68}
\eqalign{
    &\la_1=\frac16(-U-4\sqrt{\rho}),\qquad \la_2=\frac16(-U+4\sqrt{\rho}),\\
    &\la_3=\la_4=\frac16(U-4V).}
\end{equation}
As we see, two first roots coincide with the dispersionless Riemann invariants (\ref{eq63}).
The last two roots correspond to the boundary of the Whitham oscillatory region with the
smooth non-oscillatory region and they permit one to evaluate the velocity $V$ (and, hence, wavenumber) of the
modulated wave with vanishing amplitude at this boundary (see \cite{gp-74}).

In a nonlinear modulated density wave the four zeroes $\la_i$, $i=1,\,2,\,3,\,4,$ of the equation $P_0(\la)=0$
play the role of the Riemann invariants and the Whitham equations for them can be obtained by a simple
modification of the method suggested in \cite{kamch-90b,kamch-94}. From (\ref{eq21a}) and (\ref{eq25}) one can find the identity
\begin{equation}\label{eq69}
\eqalign{
    &\frac{\prt}{\prt t}\left(\frac{G_1}{g_1}\right)-\frac{\prt}{\prt x}\left(\frac{B_1}{g_1}\right)\\
    &=\frac1{g_1^2}\left[(B_2H_3-G_2C_3)g_1+(G_1C_3-B_1H_3)g_2+(B_1G_2-G_1B_2)h_3\right],}
\end{equation}
and similar identities can be obtained for other ratios of the matrix elements. In case of single-phase periodic
solution, the right-hand side of this identity vanishes as it follows from (\ref{eq28}) and (\ref{eq29}).
Hence, averaging of (\ref{eq69}) over space intervals much less than a characteristic length of change of the
slowly varying parameters yields the generating function of the averaged conservation laws
\begin{equation}\label{eq70}
    \frac{\prt}{\prt t}\left\langle\frac{G_1}{\widetilde{g}_1}\right\rangle
    -\frac{\prt}{\prt x}\left\langle\frac{B_1}{\widetilde{g}_1}\right\rangle=0,
\end{equation}
where we have made the replacement
\begin{equation}\label{eq71}
    g_1\mapsto \widetilde{g}_1=\frac{g_1}{\sqrt{P_0(\la)}}
\end{equation}
to take into account the condition that during slow evolution the identity
\begin{equation}\label{eq72}
    P_0(\la)=-f_1^2(\la)-\frac49\,g_1(\la)h_1(\la)
\end{equation}
must hold (see \cite{kamch-2000}). This identity can be easily checked for the density waves with $I_4=0$.
Hence, the imposed condition (\ref{eq72}) means that the polarization excitations are not excited during
slow evolution of the nonlinear modulated wave. Averaging can be performed according to the rule
\begin{equation}\label{eq72a}
    \left\langle\frac1{\la_i-\mu(\xi)}\right\rangle=\frac1L\int_0^L\frac{dx}{\la_i-\mu(x)}=
    \frac1L\oint\frac{d\mu}{2(\la_i-\mu)\sqrt{P_0(\mu)}}=-\frac2L\frac{\prt L}{\prt \la_i},
\end{equation}
where
\begin{equation}\label{eq72b}
    L=\oint\frac{d\mu}{3\sqrt{-P_0(\mu)}}=\frac{2K(m)}{3\sqrt{(\la_3-\la_1)(\la_4-\la_2)}},
\end{equation}
$K(m)$ is the complete elliptic integral of the first kind and
\begin{equation}\label{eq73c}
    m=\frac{\rho_2-\rho_1}{\rho_3-\rho_1}=\frac{(\la_2-\la_1)(\la_4-\la_3)}{(\la_3-\la_1)(\la_4-\la_2)}.
\end{equation}
Differentiation of the slowly varying
functions $\la_i(x,t)$ in (\ref{eq70}) and taking the limits $\la\to\la_i$ yields the Whitham equations
\begin{equation*}
    \left\langle\frac1{\la_i-\mu}\right\rangle\frac{\prt\la_i}{\prt t}-\left(\frac32-V
    \left\langle\frac1{\la_i-\mu}\right\rangle\right)\frac{\prt\la_i}{\prt x}=0,\qquad i=1,2,3,4,
\end{equation*}
or
\begin{equation}\label{eq73}
    \frac{\prt\la_i}{\prt t}+v_i\frac{\prt\la_i}{\prt x}=0, \qquad i=1,2,3,4,
\end{equation}
where
\begin{equation}\label{eq74}
    v_i=V-\frac32{\left\langle\frac1{\la_i-\mu}\right\rangle}^{-1}=V+\frac34\frac{L}{\prt_i L},\qquad \prt_i=\frac{\prt}{\prt\la_i},
\end{equation}
are the Whitham velocities. Taking into account (\ref{eq52}), we can present (\ref{eq74}) in the general form
\begin{equation}\label{eq75}
    v_i=\left(1-\frac{L}{\prt_i L}\prt_i\right)V,\qquad i=1,2,3,4.
\end{equation}
As is known, it can be obtained from the conservation law of the number of waves (see \cite{whitham})
\begin{equation}\label{eq76}
    k_t+(kV)_x=0,
\end{equation}
where $k=2\pi/L$ is a local wave vector, provided both $k$ and the phase velocity $V$ are expressed in term of the
Riemann invariants $\la_i$ as that occurs in the finite-gap integration method, and there is no any additional restriction imposed
on them. Just such a situation is realized in our case of density waves with $I_4=0$ when the periodic solution is
parameterized by four constants $\chi,\,V,\,S,\,I_3$ equivalent to the set of the Riemann invariants $\la_i$, $i=1,2,3,4$.
It is worth noticing that the Whitham equations coincide in this case with those for a periodic solution
of one-component NLS equation obtained in \cite{FL-86,pavlov-87} what is physically natural since when a relative
motion of two components of the condensate is absent then they become dynamically indistinguishable and their
motion can be described by evolution of the total density $\rho$ and the in-phase velocity $2U$.

Now let us turn to the opposite limit of purely polarization wave when the total density is constant and only
a relative motion of the components is effective.

\subsection{Whitham equations for polarization waves}

The polarization wave solution is described by the formulae (\ref{eq14}) and (\ref{eq19}) which can be written in the form
\begin{eqnarray}
    \cos\theta=\cos\beta\cos\gamma+\sin\beta\sin\gamma\cos[k(x-Vt)],\label{eq77}\\
    U=2V-\frac{k}2\left[\frac{\cos\gamma+\cos\beta}{1+\cos\theta}+\frac{\cos\gamma-\cos\beta}{1-\cos\theta}\right],\label{eq78}\\
    v=\frac{k}2\left[\frac{\cos\gamma+\cos\beta}{1+\cos\theta}-\frac{\cos\gamma-\cos\beta}{1-\cos\theta}\right],\label{eq80}
\end{eqnarray}
where the condition that the total flux $j=\rho_+v_++\rho_-v_-$ vanishes yields the relation
\begin{equation}\label{eq81}
    V=\frac12k\cos\gamma
\end{equation}
due to which the equation (\ref{eq4}) with $\rho=\mathrm{const}$ is fulfilled identically. Substitution of (\ref{eq77})--(\ref{eq80})
into (\ref{eq5}) and (\ref{eq7}) yields
\begin{equation}\label{eq82}
    U_t+(VU-V^2+k^2/4)_x=0,\qquad v_t+(Vv)_x=0.
\end{equation}
For strictly periodic waves these equations reduce to $U_t+VU_x=0$ and $v_t+Vv_x=0$ with $V=\mathrm{const}$ in
agreement with $U=U(x-Vt)$ and $v=v(x-Vt)$ in such a solution of the Manakov system. Averaging can be performed
with the use of the easily proven formulae
\begin{equation}\label{eq83}
\eqalign{
    \left\langle\frac1{1\pm\cos\theta}\right\rangle=\frac1{|\cos\gamma\pm\cos\beta|},\\
    \left\langle\frac1{(1\pm\cos\theta)^2}\right\rangle=\frac{1\pm\cos\beta\cos\gamma}{|\cos\gamma\pm\cos\beta|^3}.}
\end{equation}
As a result we get
\begin{equation}\label{eq84}
    \langle U\rangle=\left\{
    \begin{array}{c}
    2V,\quad \cos\beta>\cos\gamma,\\
    2V-k,\quad \cos\beta>\cos\gamma,
    \end{array}\right.
    \qquad
    \langle v\rangle=\left\{
    \begin{array}{c}
    k,\quad \cos\beta>\cos\gamma,\\
    0,\quad \cos\beta>\cos\gamma,
    \end{array}\right.
\end{equation}
and the averaged equations (\ref{eq82}) reduce to
\begin{equation}\label{eq85}
    k_t+(Vk)_x=0,\qquad V_t+VV_x+\frac14kk_x=0,
\end{equation}
where the first equation is the Whitham equation for conservation of number of waves which is implied in the Whitham
theory of modulations. These equations coincide with the compressible gas dynamics equations for the polytropic flow
with the ratio of specific heats equal to 3. Well-known transformation to
the Riemann invariant form yields
\begin{equation}\label{eq86}
    r_{1,t}+r_1r_{1,x}=0,\qquad r_{2,t}+r_2r_{2,x}=0,
\end{equation}
where Riemann invariants are defined by the formulae
\begin{equation}\label{eq87}
    r_1=V+k/2=\frac12k(\cos\gamma+1),\qquad r_2=V-k/2=\frac12k(\cos\gamma-1),
\end{equation}
so that the physical variables are expressed as
\begin{equation}\label{eq88}
    V=\frac{k}2\cos\gamma=\frac12(r_1+r_2),\qquad k=r_1-r_2.
\end{equation}
It is easy to see that $r_1>0$ and $r_2<0$.

Equation (\ref{eq77}) gives $\langle\cos\theta\rangle=\cos\beta\cos\gamma$ and averaging of (\ref{eq6}) yields
\begin{equation*}
    (\cos\beta\cos\gamma)_t-\frac12(k\cos\beta\sin^2\gamma)_x=0
\end{equation*}
or, after simple manipulations with the use of (\ref{eq81}) and (\ref{eq87}),
\begin{equation}\label{eq89}
    (\cos\beta)_t+\frac2{r_1+r_2}(r_1r_2\cos\beta)_x=0.
\end{equation}
This equation can be easily transformed to the diagonal form by means of introduction
of the third Riemann invariant according to
\begin{equation}\label{eq90}
    r_3=\left(\frac{r_1r_2}{r_1-r_2}\right)^2\cos\beta=\frac{k^2}{16}\sin^4\gamma\cos\beta.
\end{equation}
As a result we obtain
\begin{equation}\label{eq91}
    r_{3,t}+\frac{2r_1r_2}{r_1+r_2}\,r_{3,x}=0.
\end{equation}

Thus, we have transformed the modulation equations for a polarization wave to the Riemann diagonal form
(\ref{eq86}), (\ref{eq91}).
The initial value problem can be easily solved due to this reduction: if the initial data for $k(x,0)$,
$V(x,0)$, and $\beta(x,0)$ are given, then the initial data for $r_1(x,0),\,r_2(x,0),\,r_3(x,0)$ are also known.
Then well-known solutions of two independent Hopf equations (\ref{eq86}) yield $r_1(x,t)$ and $r_2(x,t)$ for
$t>0$. At last, a linear with respect to $r_3$ equation (\ref{eq91}) with known variable coefficients
can be solved by standard methods and that yields the complete solution of the Cauchy
problem. Applications of this method to concrete physical problems will be given elsewhere.

\section{Conclusion}

We have shown that the periodic solution of the Manakov system obtained by a direct method in
\cite{kamch-13} actually coincides with the solution obtained by the finite-gap integration method in
\cite{wright-13}. The method of Wright has the advantage that it provides the parametrization
of the solution suitable for derivation of the Whitham modulation equations for the density
waves with in-phase motion of the components. However, this parametrization in terms of the
branching points of the spectral curve corresponding to the solution is not enough for description
of the polarization wave. In Wright's method this manifests itself in the fact that the coefficients of
the spectral curve depend of the sum $I_1+I_2$ of the integrals whereas the polarization wave is
parameterized by the values of $I_1$ and $I_2$ separately. Consequently, the modulation equations
for the polarization wave have been derived by the original Whitham method of averaging of the
conservation laws without use of the complete integrability of the Manakov system.
In spite of that, the resulting equations can be transformed to the Riemann diagonal form,
as it would be naturally to expect for the completely integrable equations. The results obtained
provide the basis for applications of the theory to the problems of nonlinear dynamics of Bose-Einstein
condensates related with recent experiments \cite{hamner-11,hoefer-11,hamner-13}.

\setcounter{equation}{0}

\renewcommand{\theequation}{A.\arabic{equation}}


\appendix
\section*{Appendix A}
\setcounter{section}{1}

The systems (\ref{eq21}) have three basis solutions which we denote as columns
$\Psi^{(i)}=(\Psi^{(i)}_j)^T$, $i,j=1,2,3.$ They constitute the fundamental representation of the
group $SU(3)$. The traceless matrix $\mathbb{W}=(W_{ij})$ is expressed in terms of eight independent variables
constituting the irreducible adjoint representation $\mathbf{8}$ of $SU(3)$ (see \cite{shin-2003})
\begin{equation}\label{eqA-1}
   \mathbb{W}=\left(
                \begin{array}{ccc}
                  (\mathcal{F}_{123}-\mathcal{F}_{132})/3 & -\mathcal{F}_{113}/2 & \mathcal{F}_{112}/2 \\
                  \mathcal{F}_{223}/2 & -(2\mathcal{F}_{123}+\mathcal{F}_{132})/3 & \mathcal{F}_{122} \\
                  \mathcal{F}_{233} & -\mathcal{F}_{133} & (\mathcal{F}_{123}+2\mathcal{F}_{132})/3 \\
                \end{array}
              \right)
\end{equation}
where among 27 elements
\begin{equation}\label{eqA-2}
    \mathcal{F}_{ijk}=\Psi^{(1)}_i\Psi^{(2)}_j\Psi^{(3)}_k+\Psi^{(1)}_j\Psi^{(2)}_i\Psi^{(3)}_k
    -\Psi^{(1)}_k\Psi^{(2)}_j\Psi^{(3)}_i-\Psi^{(1)}_k\Psi^{(2)}_i\Psi^{(3)}_j
\end{equation}
only 8 are independent and they are chosen as $\mathcal{F}_{112},\,\mathcal{F}_{113},\,\mathcal{F}_{122},\,\mathcal{F}_{123},\,\mathcal{F}_{132},
\,\mathcal{F}_{133},\,\mathcal{F}_{223},\,\mathcal{F}_{233}.$
One can easily check that the matrix (\ref{eqA-1}) satisfies (\ref{eq25}), if $\Psi^{(i)}_j$
satisfy (\ref{eq21}).

\appendix

\section*{Appendix B}
\setcounter{section}{2}

We shall follow here the exposition given in \cite{dkn-85}.

Let us present the fundamental solutions of (\ref{eq21})  as a matrix
\begin{equation}\label{eqB-1}
    \mathbb{F}(x,t,\la)=(\mathbb{F}_{ij})=((\Psi^{(i)}_j)
\end{equation}
normalized by the initial data condition
\begin{equation}\label{eqB-2}
    \mathbb{F}(0,0,\la)=\mathbb{I}.
\end{equation}
It is easy to check that if $\mathbb{W}(x,t,\la)$ is the solution of (\ref{eq25}),
then the product of two matrices
\begin{equation}\label{eqB-3}
    \widetilde{\mathbb{F}}=\mathbb{W}(x,t,\la)\cdot\mathbb{F}(x,t,\la)
\end{equation}
is also the solution of (\ref{eq21}). But any solution $\widetilde{\mathbb{F}}$
is a linear combination of the fundamental solutions
\begin{equation}\label{eqB-4}
    \widetilde{\mathbb{F}}(x,t,\la)=\mathbb{F}(x,t,\la)\cdot\mathbb{G}(\la),
\end{equation}
where the matrix $\mathbb{G}$ can be determined from the normalization
condition (\ref{eqB-2}) by means of comparison of the expression (by (\ref{eqB-4}))
$$
\widetilde{\mathbb{F}}(0,0,\la)=\mathbb{I}\cdot\mathbb{G}(\la)=\mathbb{G}(\la)
$$
with (by (\ref{eqB-3}))
$$
\widetilde{\mathbb{F}}(0,0,\la)=\mathbb{W}(0,0,\la)\cdot\mathbb{F}(0,0,\la)=\mathbb{W}(0,0,\la)\cdot\mathbb{I}=\mathbb{W}(0,0,\la),
$$
that is $\mathbb{G}=\mathbb{W}(0,0,\la)$ and hence $\mathbb{F}(x,t,\la)\cdot\mathbb{W}(0,0,\la)
=\mathbb{W}(x,t,\la)\cdot\mathbb{F}(x,t,\la)$
or
\begin{equation}\label{eqB-5}
    \mathbb{W}(x,t,\la)=\mathbb{F}(x,t,\la)\cdot\mathbb{W}(0,0,\la)\cdot\mathbb{F}^{-1}(x,t,\la).
\end{equation}
Consequently, we get
\begin{equation}\label{eqB-6}
    iw\mathbb{I}-\mathbb{W}(x,t,\la)=\mathbb{F}(x,t,\la)\cdot[iw\mathbb{I}-\mathbb{W}(0,0,\la)]\cdot\mathbb{F}^{-1}(x,t,\la)
\end{equation}
and taking determinant of this identity we arrive at the statement of the theorem.

\end{document}